\documentclass{elsart}
\usepackage{graphicx,harvard,amssymb}

\usepackage[T1]{fontenc}
\usepackage[latin1]{inputenc}
\usepackage{graphics}

\makeatletter

\def\bibcode#1{(\texttt{#1})}

\def\url#1{{\ttfamily\def\/{/\discretionary{}{}{}}#1}}

\journal{New Astronomy}

\begin{document}
\begin{frontmatter}
\title{Variability in GRB Afterglows and GRB 021004}
\author[all]{Ehud Nakar\thanksref{en}},
\author[all]{Tsvi Piran\thanksref{tp}},
\author[second]{Jonathan Granot \thanksref{jg}}
\thanks[en]{E-mail: udini@phys.huji.ac.il }
\thanks[tp]{E-mail: tsvi@huji.ac.il }
\thanks[jg]{E-mail: granot@ias.edu}
\address[all]{Racah Institute for Physics, The Hebrew
University, Jerusalem, 91904, ISRAEL}
\address[second]{Institute for Advanced Studies, Princeton, NJ
08540, USA}
\begin{abstract}
We present  general analytic expressions for GRB afterglow light
curves arising from a variable external density profile and/or a
variable energy in the blast wave. The former could arise from a
clumpy ISM or a variable stellar wind; The latter could arise from
refreshed shocks or from an angular dependent jet structure
(patchy shell). Both scenarios would lead to a variable light
curve. Our formalism enables us to invert the observed light
curve and obtain possible density or  energy profiles. The optical
afterglow of GRB 021004 was detected 537 seconds AB (after the
burst) (Fox et al. 2002). Extensive follow up observations
revealed a significant temporal variability. We apply our
formalism to the R-band light curve of GRB 021004 and we find
that several models provide a good fit to the data. We consider
the patchy shell model with $p=2.2$ as the most likely
explanation. According to this model our line of sight was
towards a ``cold spot" that has lead to a relativity low
$\gamma$-ray flux and an initially weak afterglow (while the
X-ray afterglow flux after a day was above average). Observations
above the cooling frequency, $\nu_c$, could provide the best way
to distinguish between our different models.
\end{abstract}
\end{frontmatter}

\section{Introduction}

The behavior of gamma-ray burst (GRB) afterglows is well known for
a spherical shell propagating into a constant density
inter-stellar medium (ISM) or into a circum-burst wind with a
regularly decreasing density. Sari, Piran \& Narayan (1998,
hereafter SPN98) have presented a simple analytic model for the
ISM case, assuming synchrotron emission from an adiabatic
relativistic blast wave. Chevalier \& Li (1999) generalized this
model for a circum-burst wind density profile. In both cases the
flux shows a spectral and temporal segmented power law behavior,
\( F_{\nu }\propto t^{\alpha }{\nu }^{\beta } \). The indices \(
\alpha  \) and \( \beta  \) change when a spectral break frequency
(the cooling frequency, \( \nu _{c} \), the synchrotron frequency,
\( \nu _{m} \), or the self absorption frequency, \( \nu _{sa} \))
passes through the observed band. The values of the spectral and
temporal indices depends on the cooling regime (fast or slow) and
on the ordering of \( \nu \) relative to \( {\nu }_{sa} \), \(
{\nu }_{c} \) and \( {\nu }_{m} \). Most GRB afterglows display a
smooth power law decay.

In several cases the observed afterglow light curves have shown
deviations from a smooth power law. The most prominent case is the
recent GRB 021004 whose optical counterpart was observed at a very
early time,  $537\;$ sec after the trigger (Fox et al. 2002).
Following observations at short intervals showed fluctuations
around a power law decay. We develop here the general theory for
GRB afterglows when the relativistic blast wave encounters a
variable external density or its energy (per unit solid angle)
varies with time. Such variations in energy could arise due to
refreshed shocks, when initially slower moving matter encounter
the blast waves after it has slowed down (Kumar \& Piran, 2000a), or
due to angular variability within the relativistic jet (Kumar \&
Piran, 2000b). Both variation in the density or in the energy can
reproduce a variable light curve and in particular the observed
R-band light curve of GRB 021004. However, as we argue latter,
there are some weak indications that a variable energy model that
arises from a patchy shell structure (random angular fluctuations
in the jet) seems to give the best fit to all the available data.
If correct this interpretation implies that the electron power
law index is $p \approx 2.2$, a suggestion that might be
confirmed with a more detailed multi-wavelength spectrum.

\section{Theory}

We generalize the results of SPN98 to a time dependent energy and
a spatially varying external density. We first outline the
general model and then investigate  two specific cases. Following
SPN98 we assume that the dominant radiation process is
synchrotron emission. In our model, the mass in the blast-wave at
radius \( R \) is taken to be the integrated external mass up to
this radius, and we assume that all this mass is radiating. The
internal energy density of the emitting matter at radius \( R \)
is taken from the shock jump conditions, which depend only on \(
\gamma (R) \) and \( n(R) \). These approximations are valid as
long as the external density and the energy in the blast-wave do
not vary too rapidly.  For example a large density jump can
produce a reverse shock while a sharp density drop may initiate a
rarefaction wave. The accuracy of this model decreases as the
variations in the density and the energy become more rapid.

A few hundred seconds after the GRB the relativistic ejecta
decelerates, driving a strong relativistic shock into the ambient
medium. As radiative losses become negligible the flow settles
into the adiabatic  self-similar Blandford-McKee (1976) blast
wave solution. Energy conversion takes place within the shock that
propagates into the external medium. The  energy  equation reads:
\begin{equation}\label{E}
E(t)=A\gamma^{2}[R(t)]M[R(t)]c^{2}\ ,
\end{equation}
where  \( E \) is the isotropic equivalent energy and \( A \) is a
constant of order unity whose exact value depends on the density
profile behind the shock (e.g. for an external density
$n(r)\propto r^{-k}$, $A=4(3-k)/(17-4k)$; Blandford \& McKee 1976). In the
following we use \( A=1 \). In equation (\ref{E}), \( M(R) \) is the mass
of the blast-wave, i.e. the integrated external mass up to a radius \( R \),
\begin{equation}
\label{Mass eq} M(R)\equiv 4\pi \int _{0}^{R}n(r)r^{2}dr\ .
\end{equation}

The observed time, $t$, is related to \( R \) and \( \gamma  \)
through two effects. First, the observed time of a photon emitted
on the line of sight at a radius \( R \) is \( t_{\rm
los}=(1/4c)\int ^{R}_{0}\gamma ^{-2}dr \). Second, photons
emitted at different angles at the same radius \( R \) are
observed during an interval of \(\sim R/2c\gamma ^{2} \).
Following SPN98 we estimate the observed time interval during
which most of the emission emitted at radius \( R \) is received
as \( t_{\rm ang}\approx R/4c\gamma ^{2} \). Therefore:
\begin{equation} \label{eq tobs} t={1\over 4c}\left(
{R\over\gamma^2}+\int^{R}_{0}{dr\over\gamma^{2}}\right)\ ,
\end{equation}

For a constant density ISM \( t_{\rm ang}=4t_{\rm los} \) and \(
t=5R/16c\gamma ^{2}\approx R/4\gamma ^{2}c \). Of course, this
treatment of the angular effects is only approximate \footnote{
See the Appendix for an extended  discussion of the angular
smoothing effect and  Nakar \& Piran (2003) for a  solution that
takes a full account of this effect}. In most cases angular
spreading will smooth out any variability on time scales shorter
than \( R/4c\gamma ^{2} \).

We further assume that the electron energy distribution is a power
law with an index \( p \), and that the magnetic field and the
electrons hold fractions \( {\epsilon }_{B} \) and \( {\epsilon
}_{e} \), respectively, of the internal energy. Now, taking \( \nu
_{m}(\gamma ,n,{\epsilon }_{B},{\epsilon }_{e}) \), \( \nu
_{c}(\gamma ,n,{\epsilon }_{B},t) \) and \( F_{{\nu
},max}(M,\gamma ,n,{\epsilon }_{B}) \) from SPN98 and the
equations above we obtain:

\begin{equation}\label{nu_m}
\nu _{m}=5\cdot 10^{12}E_{52}^{2}M_{29}^{-2}n_{0}^{1/2}{\epsilon
}_{B_{-2}}^{1/2}{\epsilon }_{e_{-1}}^{2}\;{\rm Hz}\ ,
\end{equation}
\begin{equation}
\nu _{c}=3\cdot
10^{14}E_{52}^{-2}M_{29}^{2}n_{0}^{-3/2}t_{d}^{-2}{\epsilon
}_{B_{-2}}^{-3/2}\;{\rm Hz}\ ,
\end{equation}
\begin{equation}\label{F_nu_max}
F_{{\nu },{\rm max}}=7\,E_{52}n_{0}^{1/2}{\epsilon
}_{B_{-2}}^{1/2}D^{-2}_{28}\;{\rm mJy}\ ,
\end{equation}
where \( Q_{x} \) denotes the value of the quantity Q in units of
\( 10^{x} \) (c.g.s), \( t_{d} \) is the observed time in days,
$D$ is the distance to the GRB, and for simplicity we do not
include cosmological effects throughout the paper. The above
equations readily provide expressions for the flux density at
different frequencies
\begin{equation} \label{eq Fnu Generic} F_{\nu }\propto \left\{ \begin{array}{c}
\nu^{(1-p)/2}E^{p}M^{1-p}n^{(1+p)/4}{\epsilon }_{B}^{(1+p)/4}{\epsilon }_{e}^{(p-1)}
\qquad \nu _{m}<\nu <\nu _{c}\\
\nu^{-p/2}E^{p-1}M^{2-p}n^{(p-2)/4}t^{-1}{\epsilon
}_{B}^{(p-2)/4}{\epsilon }_{e}^{(p-1)}\qquad \nu _{c}<\nu
\end{array}\right.\ .
\end{equation}
We concentrate on the above two power law segments, since they are
usually expected to be the most relevant for the optical light
curve. Similar expressions for other power law segments of the
spectrum may be derived similarly.

These are the generic expressions for a varying energy and a
varying external density profile. In addition to the explicit
dependence on \( t \) in Eq. \ref{eq Fnu Generic} there is an
implicit dependence through \( E(t) \), \( M[R(t)] \) and \(
n[R(t)] \). For an ISM or wind, $M[R(t)]$ and $n[R(t)]$ have
simple analytic forms and Eq. \ref{eq Fnu Generic} reduces to the
expressions of SPN98 and Chevalier \& Li (1999).

For \( p\approx 2 \), \( F_{\nu >\nu _{c}} \) is only weakly
dependent on \( M \) and \( n \), while the dependence on \( E \)
is roughly linear (note that \( F_{\nu } \) depends on \( E \)
also implicitly through \( R \) that appears in \( M(R) \)). This
feature enables us to  distinguish between energy dominated
fluctuations and density dominated fluctuations  in the
afterglow light curve, when there are measurements both above
and below the cooling frequency, $\nu_c$.

In reality, it is unlikely that both variations (in $E$ and in
$n$) will be important in a given burst (since this would require a
coincidence). Therefore, we shall consider below, in some
detail, the cases where one of these quantities is constant while
the other one varies. Moreover, the information in a single band
light curve  (or more accurately, from a single power law segment
of the spectrum) is insufficient to determine both profiles. For
any given set of density and energy profiles the light curve can
be easily calculated. However, these profiles are not at hand.
The observable is the light curve and these profiles are unknown
variables. It is necessary to make some assumption for one of the
profiles in order to deduce the other (for example, to assume a
constant energy or a constant density).

\subsection{A Variable External Density}

Consider, first, the case where the dominant variations are in the density profile
while the energy is constant. Eqs. \ref{Mass eq}, \ref{eq tobs} and \ref{eq Fnu Generic}
reduce to: \begin{equation}
\label{eq Fnu Econst}
F_{{\nu }}\propto \left\{ \begin{array}{c}
M^{1-p}n^{(1+p)/4}\; \quad \nu _{m}<\nu <\nu _{c}\\
M^{2-p}n^{(p-2)/4}t^{-1}\qquad \nu _{c}<\nu
\end{array}\right. \ ,
\end{equation}

\begin{equation}
\label{eq t Econst} t=\frac{c}{4E}\left( MR+\int
^{R}_{0}Mdr\right) \ ,
\end{equation}
\begin{equation}
\label{eq M Econst} M=4\pi m_{p}\int ^{R}_{0}r^{2}n(r)dr \ .
\end{equation}

For a given \( F_{\nu }(t) \) we solve Eqs. \ref{eq Fnu
Econst}-\ref{eq M Econst} for \( R(t) \), \( n[R(t)] \) and \(
M[R(t)] \) with \( p \) as a free parameter. The integral
dependence of \( M[n(r)] \) in Eqs. \ref{eq t Econst} and \ref{eq
M Econst} makes it difficult to invert these equations
analytically for an arbitrary density profile (an exact numerical
solution is always possible). However, an approximate analytic
solution can be obtained if the density profile varies slowly
(note that as discussed earlier, when the density varies rapidly
our whole approach is less accurate).

As \( M \) grows monotonically with \( R \), \( t_{\rm ang} \) is
always larger than \( t_{\rm los} \) and we can approximate \(
t\approx t_{ang} \). Taking the time derivative of Eq. \ref{eq Fnu
Econst} for \( \nu _{m}<\nu <\nu _{c} \) and using Eq. \ref{eq M
Econst} we obtain:

\begin{equation}
{d\ln F_\nu\over d\ln t} = \delta_m  + {(1+p)\over 4}{d\ln n\over
d\ln t} \quad {\rm for} \quad \nu _{m}<\nu <\nu _{c}\ ,
\end{equation}
where \( \delta _{m}\equiv (1-p)/[1+\overline{n}/(3n)] \) and \(
\overline{n}(R)=M/(4/3)\pi m_{p}R^{3} \) is the average initial
density inside a sphere of radius \( R \). If \( \delta _{m} \)
varies slowly with time we derive:

\begin{equation}
\label{eq Fnu<nuc analytic} F_{\nu }=F_{0}(t/t_{0})^{\delta
_{m}}\: (n/n_{0})^{(1+p)/4}\quad {\rm for}\quad \nu _{m}<\nu <\nu
_{c}\ ,
\end{equation}
where \( F_{0} \) and \( n_{0} \) are the flux and density at some
given time \( t_{0} \). As long as \( n>\overline{n} \), \( \delta
_{m} \) depends weakly on \( \overline{n}/n \) and its value
varies between \( 1-p \) and \( 0.75(1-p) \). When \( n\ll
\overline{n} \), \( \delta _{m}\rightarrow 0 \) and \( F_{\nu } \)
depends on \( t \) only implicitly via \( n \).

A similar derivation for \( \nu _{c}<\nu  \) results in:

\begin{equation}
\label{eq Fnu>nuc analytic} F_{\nu }=F_{0}(t/t_{0})^{\delta
_{c}}\: (n/n_{0})^{(p-2)/4}\quad {\rm for}\quad \nu _{c}<\nu \ ,
\end{equation}
where \( \delta _{c}=(2-p)/[1+(1/3)\overline{n}/n]-1 \). The
explicit dependence on \( n \) is negligible, \( n^{(p-2)/4} \),
and the variations in \( \overline{n}/n \) yield \( 1-p<\delta
_{c}<-1 \). The variations in \( \delta _{c} \) could be measured
if \( p \) is large and \( \overline{n}\gg n \). However, in this
limit the density changes very rapidly, so that our formalism may
not hold. Both Eqs. \ref{eq Fnu<nuc analytic} and \ref{eq Fnu>nuc
analytic} contain the wind solution (with \( 3n=\overline{n} \)
and \( n\propto R^{-2}\propto t^{-1} \)) and the ISM solution
(with \( n=\overline{n} \) ).

\subsection{A Varying Energy}

Consider now the afterglow when the energy in the emitting region varies with
time but the density profile is regular. In the ISM case Eqs. \ref{eq Fnu Generic}
and \ref{eq tobs} are reduced to:
\begin{equation}
\label{eq Fnu ISM}
F_{\nu }\propto \left\{ \begin{array}{c}
E^{p}R^{3(p-1)}\qquad \nu _{m}<\nu <\nu _{c}\\
E^{p-1}R^{3(2-p)}t^{-1}\quad \: \nu _{c}<\nu
\end{array}\right. \quad {\rm (ISM)}\ ,
\end{equation}
\begin{equation}
\label{eq t nconst} t=\frac{\pi nm_{p}c}{3}\left(
\frac{R^{4}}{E}+\int ^{R}_{0}\frac{r^{3}}{E}dr\right) \ .
\end{equation}
 In the wind case ( \( n=A_{w}R^{-2} \)) these equations become:\begin{equation}
\label{eq Fnu WIND}
F_{\nu }\propto \left\{ \begin{array}{c}
E^{p}R^{(1-3p)/2}\qquad \nu _{m}<\nu <\nu _{c}\\
E^{p-1}R^{1.5(2-p)}t^{-1}\qquad \nu _{c}<\nu
\end{array}\right. \quad {\rm (wind)}\ ,
\end{equation}
\begin{equation}
t=\pi A_{w}m_{p}c\left( \frac{R^{2}}{E}+\int
^{R}_{0}\frac{r}{E}dr\right)\ .
\end{equation}
Again, these equations can be solved numerically for a given \(
F_{\nu }(t) \). Note that in this case the condition \( t_{\rm
ang}>t_{\rm los} \) does not always hold. A sharp increase in \( E
\) would decrease \( t_{\rm ang} \) without affecting \( t_{\rm
los} \). However, if the energy profile is not too steep, the
condition \( t_{\rm ang}>t_{\rm los} \) does hold, and we can
approximate \( t \) by \( t_{\rm ang} \). In this case, Eqs.
\ref{eq Fnu ISM} and \ref{eq Fnu WIND} reduce to the well known
ISM and wind equations for a constant energy, where \( E \) is
replaced by \( E(t) \).

Two different phenomena could cause energy variations in the
emitting region: refreshed shocks and initial energy
inhomogeneities in the jet. Refreshed shocks (Kumar \& Piran
2000a) are produced by massive and slow shells, ejected late in
the GRB, that take over the blast-wave at late times, when the
blast-wave has decelerated. These shells bring new energy into the
blast-wave. The collision produces a refreshed forward shock
propagating into the blast-wave and a reverse shock propagating
into the slower shell. After these shocks cross the shells the
blast-wave  relaxes back to a Blandford \& McKee (1976)
self-similar solution  with a larger total energy (Since the
mass of the blast wave is dominated by the swept circum-burst
material, we neglect the mass of the inner shell). At this stage
the observed flux is similar to the one emitted by a constant
energy blast wave with the new  and larger energy. Refreshed
shocks can only increase the energy. Therefore a refreshed shocks
energy profile should grow monotonically with time, most likely in a
step wise profile (each step corresponds to the arrival of a new
shell).

Initial energy inhomogeneities (the patchy shell model of Kumar \&
Piran 2000b) in the jet could be either regular or irregular
ones. During the jet evolution regions within the relativistic
flow with an angular separation larger than $\gamma^{-1}$ are
casually disconnected. Therefore, the inhomogeneities could be
smoothed only up to an angular scale of \( \gamma ^{-1} \). As \(
\gamma \) decrease the causal connected regions grow and the
initial inhomogeneities can be smoothed on angular scale of \(
\gamma ^{-1} \). Recent numerical hydrodynamical studies (Kumar
\& Granot 2002) show that at early times the initial fluctuations
remain almost unchanged, and are smoothed only at rather late
times. Additionally, due to relativistic beaming, an observer can
see only regions within an angle of \( \gamma ^{-1} \) around the
line of sight. However, regardless of the degree of hydrodynamical
smoothing of the initial fluctuations, when combined with the
relativistic beaming, the two effect cause $F_\nu(t)$ to reflect
the initial physical conditions within a solid angle of \(
\sim\gamma ^{-2}(t) \). As a consequence, the average energy in
the observed area varies with \( \gamma  \) and therefore with \(
t \). This behavior can be approximated by the solution presented
above, where \( E(t) \) is the averaged initial isotropic
equivalent energy within a solid angle of \( \gamma ^{-2} \),
$\bar E(t)$.

In the patchy shell scenario, fluctuations would appear in the
energy profile when \( \gamma ^{-1} \) increases to the typical
angular size, $\theta_{\rm fl}$, of the initial inhomogeneities.
When $\gamma^{-1}\sim\theta_{\rm fl}$ the nearest neighboring
fluctuations begin to be observed, and the amplitude of the
fluctuations in $\bar E$ (and correspondingly in $F_\nu$) are
largest, of the order of the amplitude of the individual
fluctuations, $A_{\rm fl}$. As $\gamma$ decreases below
$\theta_{\rm fl}^{-1}$, the observed number of fluctuations
becomes large, $N_{\rm fl} \sim(\gamma\theta_{\rm fl})^{-2}$,  and
the amplitude of the fluctuations in $\bar E$ decreases to $\sim
A_{\rm fl}N_{\rm fl}^{-1/2}\sim A_{\rm fl}\gamma\theta_{\rm
fl}\propto\gamma$. For $\nu>\min(\nu_m,\nu_c)$, $F_\nu$ has a
close to linear dependence on $E(t)\approx\bar E(t)$, so that the
amplitude of the fluctuations in $F_\nu$ should be similar to
those in $\bar E$, with only minor differences between the
different power law segments of the spectrum.

A single bump in the light curve can be seen for an axially
symmetric structured jet, by an observer at an angle
\( \theta_{\rm obs} \) from the jet symmetry axis,
at the time when $\gamma ^{-1}\approx\theta_{\rm obs}$.
At this time the brighter portion of the jet, near its symmetry axis
where the energy per unit solid angle is largest, becomes visible to
an observer at angle \( \theta_{\rm obs} \). Additional
bumps are more difficult to produce.

\section{The Light Curve of GRB 021004}

GRB 021004 is a faint long burst detected by Hete-2 Fregate
instrument. The burst redshift is z=2.232 (Chornock \&
Filippenko, 2002) and its isotropic equivalent energy is $6
\cdot 10^{52}$ergs (Lamb et al. 2002 and Malesani et al.
2002). An optical counterpart  was first observed  $537\;$sec
AB (after the burst) (Fox et al. 2002) at an R magnitude of
15.5. After a short power law decay,  at $t \sim 2000\;$sec, a
clear bump (about $1.5\;$mag above the power law decay) is observed.
>From this time on, frequent observations showed a fluctuating
light curve (possibly above and below a power law decay).
The inset of Fig. \ref{fig:n(R)} shows the R-band light
curve up to 5 days after the trigger.
Observations after 6 days show a steepening of the
light curve which  may be interpreted as a jet break (Malesani et
al. 2002). A break at this time implies a total energy (after
beaming corrections) of\footnote{ This value
is obtained using a redshift of  2.323 and isotropic equivalent
energy of $6 \cdot 10^{52}\;$ergs. The rest of the parameters are
similar to those  of Frail et. al. 2001.} $3 \cdot 10^{50}\;$ergs.
Chandra observed the X-ray counterpart of
GRB 021004 at $20.5\;$hr AB for a duration of $87\;$ks (Sako \&
Harrison 2002). The corresponding mean 2 - 10 keV X-ray flux
in the observer frame is $4.3 \cdot 10^{-13}\;{\rm erg}~{\rm
cm}^{-2}~{\rm sec}^{-1}$. The X-ray observations showed a power
law decay index of $-1 \pm 0.2$ and a photon index of $2.1 \pm
0.1$ which  imply an electron index $p=2.2 \pm 0.2$.

We use the two models described above  to find a varying density
profile or a varying energy profile that reproduce the light curve
of GRB 021004. We fit the R-band  light curve that has the most
detailed data. Unfortunately, the data in the other bands is not
detailed enough and the effect of reddening is  unknown so a multi
wavelength fit is impossible at this stage. We assume that the R
band is above the synchrotron frequency, $\nu_m$, and below the
cooling frequency, $\nu_c$. This assumption is marginal at
the time of the first bump (Both the transition from fast to slow
cooling and the passage of $\nu_m$ through the optical bands occur
approximately at this time). However, this assumption is certainly
valid during the later fluctuations of the light curve\footnote{
It is possible that the origin of the first bump is different from
the later fluctuations (e.g. a passage of $\nu_m$ through the R
band combined with the emission from the reverse shock, Kobayashi
\& Zhang 2002), but following Occam's razor we are looking for a
single explanation to the whole light curve.}. It has been
suggested that $\nu_c$ passes through the optical at $t\sim
1-3\;$days (Matheson et al. 2002). In this case, we expect the
fluctuations in the light curve to decrease dramatically at
$t>3\;$ days, if they are due to fluctuations in the external
density. We discuss only variability above a constant ISM density
profile. As we show latter, a reasonable fit with a background
wind profile requires an electrons' index $p<2$ (for either variable
density or variable energy), which we consider to be a not very
physical value.

\subsection{A Variable Density Profile}

Lazzati et al. (2002) suggest that the fluctuations seen in the
R-band light curve arise from  variations in  the external density
profile. They calculate numerically the resulting light curve for
a given density profile, assuming $p=2$, and show that it agrees
with the observations. We invert the observed R-band light curve,
both analytically and numerically, and derive several possible density
profiles for different values of $p$.

\begin{figure}
\resizebox*{0.9\columnwidth}{0.35\textheight}{\includegraphics{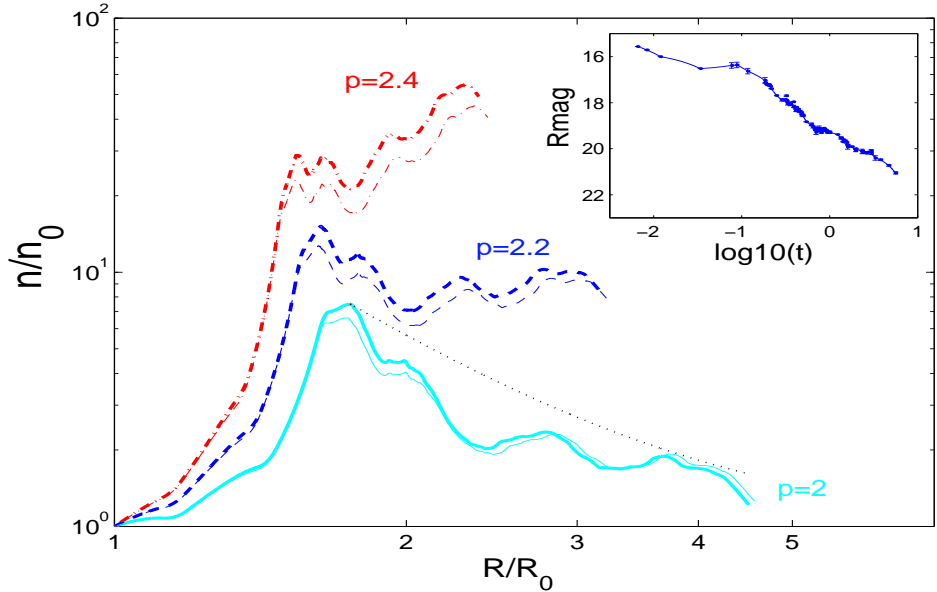}}
\caption{\label{fig:n(R)} The external density profiles, $n(R)$,
that
  reproduce the R-band light curve, for different values of electrons'
  index, $p$: $p=2.4$ (dashed-doted), $p=2.2$ (dashed line) and $p=2$
  (solid line). The thick lines are the exact numeric solution of Eqs.
\ref{eq Fnu Econst}-\ref{eq M Econst}. The thin lines are the
analytic solution of Eq. \ref{eq Fnu<nuc analytic}. The thin
dotted line depicts the expected amplitude of the density
fluctuations, $\Delta n(R)\propto R^{-5/2}$, for a random
distribution of clumps inside a uniform density background, $n_0$.
The normalization is derived assuming that the first bump is due
to a single clump. The inset on the right depicts the R-band
observed data points and the fitted light curve. The observed
R-band data points are taken from: Fox 2002, Uemura et. al.
2002, Oksanen \& Aho 2002, Rhoads et. al. 2002, Winn et.
al. 2002, Zharikov et. al. 2002, Halpern et.al. 2002a \&
2002b, Balman et. al. 2002, Cool \& Schaefer 2002,
Holland et. al. 2002a \& 2002b, Bersier et. al. 2002, Sahu
et. al. 2002, Oksanen et. al. 2003,
 Matsumoto et.al. 2002, Stanek et. al. 2002, Mirabel
et. al. 2002, Masseti et. al. 2002, Barsukova et. al. 2002
, Malesani et. al. 2002, Mirabel et. al. 2002.}
\end{figure}

We begin the fit at the first observation, $t_0=537\;$sec after
the trigger, and define \( n_{0} \) and \( R_{0} \) as the values
at this time. For simplicity, we assume a constant density up to
$R_0$ [so that $n(R<R_{0})=n_{0}$]. With this assumption the
ratios \( R/R_{0} \) and \( n/n_{0} \) do not depends on the
values of \( R_{0} \) and \( n_{0} \). Figure \ref{fig:n(R)}
depicts the density profile for a few values of the electron
power law index, \( p \). The thick lines show the exact numerical
solution of Eqs. \ref{eq Fnu Econst}-\ref{eq M Econst},
 while the thin lines show the analytic solution of Eq. \ref{eq
Fnu<nuc analytic} (In this solution the value of ${\delta}_m$ is
recalculated every time step). In order to reproduce the light
curve with \( p \ge 2.4 \), the density profile must increase
with $R$ almost monotonously. Such a density profile does not
look feasible. For \( p=2.2 \) the density increases by an order
of magnitude at \( R\approx 1.5R_{0}\sim 3\times 10^{17}\;{\rm
cm} \) and remains roughly constant at larger radii. This is
consistent with the termination shock of a stellar wind that
interacts with the ambient medium (Wijers 2001), provided that
the latter has a very high density of $\sim 10^{4-5}\;{\rm
cm}^{-3}$ in order for the radius of the wind termination shock
to agree with the afterglow shock radius inferred from the time
of the first bump. When \( p=2 \) the density profile rises by
almost an order of magnitude and then decreases, more gradually,
back to its initial value. The initial rise agrees with the one
suggested by Lazzati et al. (2002), however, Lazzati et al.
suggest a consequent decrease in the density to a factor of 5
below the initial density value followed by a second and smaller
density bump, where according to our results such a large dip in
the density is not required. The difference between the profiles
arises  mainly due to the different approximation used for the
angular smoothing effect (see the Appendix).

So far, we have assumed a spherically symmetric external density
profile, $n=n(r)$. This may occur due to a variable stellar wind,
but is not expected for an ISM. As we obtain that an underlying
constant density profile provides a better fit for GRB 0210004,
it is more natural to expect density fluctuations in the form of
clumps, rather than being spherically symmetric, in this case.
This interpretation requires $p\approx 2$ for which the density
at large radii decreases back to its value at $R_0$. As the
density profile for $p=2$ is not smooth, several density clumps
are needed. The first clump should be at $R_1\approx 1.5R_0$ and
with an over-density of factor $\sim 8$. In order to have a
similar effect as a spherical density bump, the clump must
replace all the emitting material, i.e. its size (radius),
$l_{\rm cl}$, must be large enough so that its mass is larger
than the swept up mass at that radius within an angle of
$\gamma^{-1}$ around the line of sight: $l_{\rm cl}>l_{\rm
min}=(n_0/n_{\rm cl}4\gamma^{2})^{1/3}R_1\approx 0.03R_1\approx
10^{16}\;$cm. An upper limit on the size of the clump can be put
from the fact that the bump in the light curve decays on a time
scale $\Delta t\sim t$. Since $R\propto t^{1/4}$ for an ISM, this
implies $l_{\rm cl}<l_{\rm max}=(2^{1/4}-1)R_1\approx 0.19R_1$
(Lazzati et al. (2002) obtain a similar clump size using
different considerations).

Assuming a homogeneous distribution of clumps with the same
physical size and over-density, the mean distance between
neighboring clumps is $d_{\rm cl}\sim(\pi l_{\rm cl}^2
R_1)^{1/3}\approx 4-5\times 10^{16}\;$cm, where the numerical
estimate assumes $l_{\rm cl}=l_{\rm min}$, in which case the
clumps hold roughly $5\%$ of the volume and $30\%$ of the mass
(these are lower limits as $l_{\rm cl}>l_{\rm min}$ would imply
larger filling factors). Therefore, soon after the collision with
the first clump we expect overlap between pulses from different
clumps, where the number of clumps that intersect a given shell
with a radius R and angular size  $1/\gamma$ is $N_{\rm cl}\sim
R^2 l_{\rm cl}/\gamma^2 d_{\rm cl}^3\propto R^{5}$.
Since, on average, the clumps hold a constant fraction of the
shell's mass, a single clump constitute a fraction $\propto R^{-5}$
of the  matter at this radius. The total fluctuation in the
density would be, therefore, $\propto R^{-5/2}$. This is in a
rough agreement with the fluctuations in the density profile we
have obtained for p=2 (see Fig \ref{fig:n(R)}).

\subsection{A Variable Energy Profile}

\begin{figure}
\resizebox*{0.9\columnwidth}{0.35\textheight}{\includegraphics{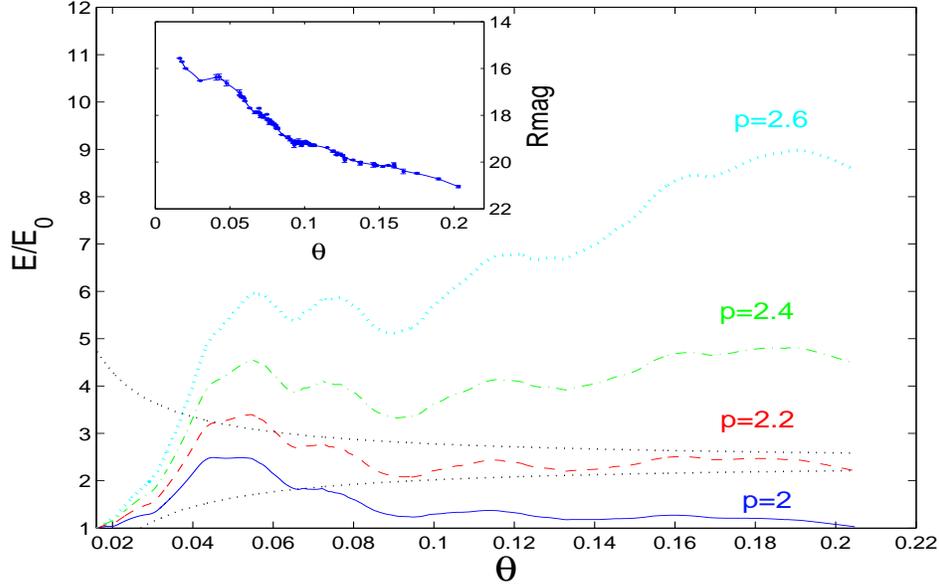}}
\caption{\label{fig E} The isotropic equivalent energy, $E$,
within an angle $\theta=1/\gamma$ around the line of sight as a
function of $\theta$, for different values of $p$: $p=2$ (solid
line), $p=2.2$ (dashed line), $p=2.4$ (dashed-dotted line) and
$p=2.6$ (bold dotted line). The curves are normalized by the value
at the first observation, $E_0$. The thin dotted lines outline the
expected fluctuations in $E$ for a patchy shell model with
fluctuations on an angular scale $\theta_{\rm fl}=0.02$ (using
$E=2.4E_0\pm 1.9E_0(\theta_{\rm fl}/\theta)$). The inset on the
left depicts the observed data points as a function of $\theta$
and the fitted curve.}
\end{figure}

We solve Eqs. \ref{eq Fnu ISM} and \ref{eq t nconst} numerically,
for a constant ISM density profile, assuming that the energy is
constant, \( E_{0} \), up to the first observation at \( t_{0}
\), and letting \( E \) vary from this point onwards. Figure
\ref{fig E} depicts the energy profile obtained for different
values of \( p \) as a function of \( \theta =1/\gamma \) (the
angular size of the observed area). An electron power law index of
\( p=2.6 \), requires an almost monotonous increase of \( E \) in
the observed region. Such a profile may arise due to refreshed
shocks. However, the continues increase in \( E \) requires a
continuous arrival of new shells, a scenario which we consider as
unlikely. The energy profiles obtained for \( p=2.2 \) and \(
p=2.4 \) could reflect  irregular patches with an initial angular
size of $\theta_{\rm fl}\approx 0.02\;$rad and an average energy
of several times \( E_{0} \). The energy fluctuations decrease
with time, as expected from a patchy shell (see Figure \ref{fig
E}). The profile obtained for \( p=2 \) shows an initial rise
followed by a gradual (and bumpy) decrease back to the initial
value. Such a profile can correspond to a line of sight is $\sim
0.04\;$rad away from a hot spot (the average energy over a large
area is \( E_{0} \)). This hot spot may be a hot patch in an
irregular jet. Alternatively,  as suggested by Lazzati et al. (2002),
this hot spot may be the core of a
jet (on the jet axis) in an axisymmetric angle dependent regular
jet\footnote{Though the wiggles in $E(\theta)$ require some
additional small amplitude variability on small angular scales on
top of an underlying smooth axisymmetric jet profile on large
angular scales.}. According to this interpretation the angular
size of the jet's core is $\theta_c\sim 0.02\;$rad, the isotropic
equivalent energy outside the core is roughly constant and its
value is $\sim 3$ times less than the core's energy.

\section{Discussion}

We have presented general expressions for the afterglow light
curve when the energy in the blast wave varies with time and for
a variable external density profile. This formalism follows and
generalizes the work of SPN98, and relates the variability in the
energy and density to the variability in the light curve. Despite
the variability in the light curve, the shape of the broad band
spectrum remains the same, with some variability in the values of
the break frequencies and flux normalization.

We have focused on the slow cooling spectrum at frequencies
$\nu>\nu_m$, and derived detailed equations for these cases, as
they seem the most relevant for the majority of observed optical
light curves. Similar equations can be easily derived for other
spectral regimes using Eqs. \ref{nu_m}-\ref{F_nu_max}. We find
that for $\nu_m<\nu<\nu_c$, variability in the light curve can be
induced both by variability in the energy or by variability in the
external density (or both). A similar behavior is expected for
$\nu<\min(\nu_m,\nu_c)$, for both slow and fast cooling. For
$\nu>\max(\nu_m,\nu_c)$ we find that a variable density hardly
induces any fluctuations in the light curve, while a variable
energy can induce significant fluctuations. We expect a similar
behavior for $\nu_c<\nu$, in the fast cooling regime.

We applied our formalism to GRB 021004, which displayed
significant deviations from a simple power law decay in its
optical (R-band) light curve. We find that several different
models may provide a reasonable fit to the observed light curve.
These include models where the variability is induced either by
density fluctuations or by energy fluctuations, where the latter
may be caused either by refreshed shocks or by a patchy angular
structure of the GRB outflow. These models vary significantly
with the value of $p$. Chandra's observations constrain the
electron's index to be $p=2.2 \pm 0.2$, but even under this
constrain many different models can produce the observed light
curve. A tighter constrain would limit the models considerably.
The following models provide a viable fit to the light curve: I) A
variable density: a) For $p=2.2$ there is an order of magnitude
rise in the density followed by a roughly constant density; b)
For $p=2$ we find a similar rise, but then the density gradually
decreases back to its initial value; II) A variable energy: a)
For $p=2.6$ refreshed shocks are required in order to explain the
energy profile; b) For $p\approx 2.2-2.4$ a patchy shell model
provides a good fit; c) For $p=2$ a hot spot (possibly the core
of an axisymmetric jet) should reside near our line of sight.

\begin{figure}
\resizebox*{0.9\columnwidth}{0.35\textheight}{\includegraphics{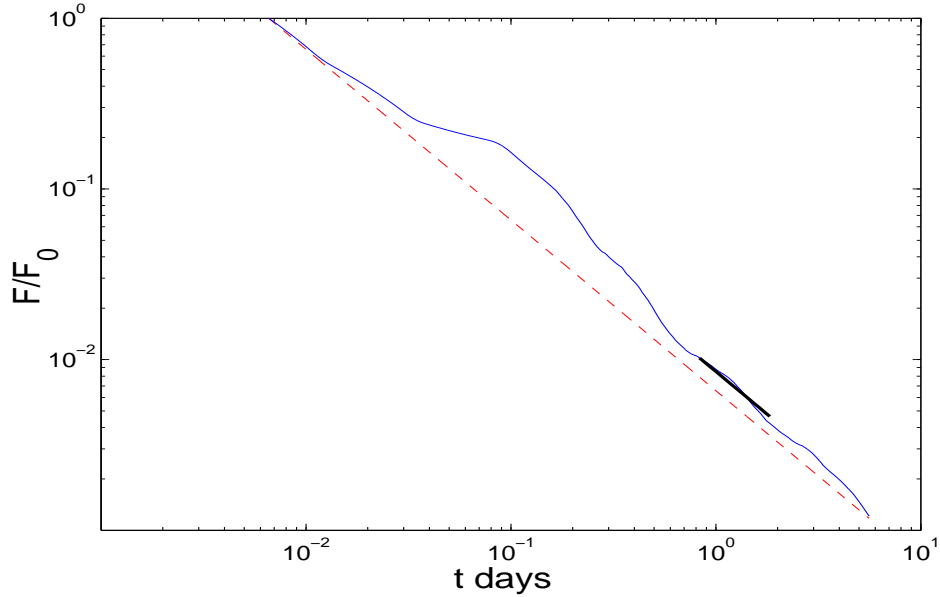}}
\caption{\label{fig:lc_above_nuc} The expected light curve
$F_{\nu}(t)/F_{\nu,0}$ for $\nu > {\nu}_c$ where $F_{\nu,0}$ is the
observed flux at $t_0=537\;$sec. The expected light curves are
calculated using (i) the energy profile that reproduces the R-band
light curve for $p=2$ (solid line) and (ii) the density profile
that reproduce the R-band light curve for $p=2$ (dashed line).
The short thick line represents Chnadra's X-ray measurement,
normalized for the expected flux at $20.5\;$hr in the varying energy
light curve (a power law decay with index $-1$ from $20.5\;$hr till
$44\;$hr after the trigger). }
\end{figure}

As any given single band light curve (which does not show a strong
variability on time scales $\Delta t\ll t$) can be reproduced by
either  density or  energy variations, it is important to find
ways to distinguish between these two models and their variants.
An independently determined value of $p$, say from the spectrum,
would have made this task easier (but still not completely
determined). Simultaneous light curves both above and below the
cooling frequency, $\nu_c$, provide the best way to differentiate
between a variable energy and a variable density: for the latter
strong variability is possible only below $\nu_c$. Figure
\ref{fig:lc_above_nuc} depicts the light curves that are
predicted above $\nu_c$, using the energy or density profiles
deduced from the R-band light curve, that is assumed to be below
$\nu_c$. Chandra obtained an X-ray light curve between 1 and 2
days (Sako et al. 2002, the thick lines in Figure
\ref{fig:lc_above_nuc}). Unfortunately, by this time the
fluctuations expected in the X-ray light curve according to the
two models are rather similar and it is hard to distinguish
between them. Still, it would be interesting to search for a
correlation between the R-band light curve and the X-ray light
curve at this time.  An earlier X-ray observation could have
enabled a clear distinction between the two models.

A variable energy model could arise either from refreshed shocks or
from angular inhomogeneity in the jet. In the refreshed socks
scenario, we expect during the collision between the two shells
an increase in the spectral slope $\beta$ (defined by
$F_\nu\propto\nu^\beta$) and a strong signal in the radio, (Kumar
\& Piran 2000a). This emission should last over $\Delta t\sim t$. A
refreshed shocks can only add energy to the blast wave the total
energy in this picture can only increase with time. In the patchy
shell model we expect random fluctuations  whose amplitude decays
with time as $1/\gamma$ (see Fig. \ref{fig E}).

Although the current observations do not enable us to determine
which one of the scenarios described above is the correct one (if
any), we feel that the patchy shell model with $p=2.2$ (which
agrees with the $p=2.2 \pm0.2$ value suggested by Chandra's
observations) is the most likely scenario.  According to this
interpretation the line of sight of GRB 021004 falls in a ``cold
spot"  where the energy is 2.5 times below the average. This
agrees with the observation of rather low $\gamma$-ray flux from
this burst. The total $\gamma$-ray energy, $E_\gamma =
3 \cdot 10^{50}\;$ergs  is within the standard deviation of the
energy distribution presented by Frail et. al (2001), but it is 1.5
times smaller than the average value. On the other hand an
extrapolation of Chandra's measured X-ray flux (Sako et al.
2002) to $11\;$hr after the burst yields
$F_x\approx 9 \cdot 10^{-13}\;{\rm ergs}~{\rm cm}^{-2}~{\rm sec}^{-1}$.
This value is 1.5 times larger than the narrowly
clustered value of $F_x$ in other bursts:
$6 \cdot 10^{-13}\;{\rm ergs}~{\rm cm}^{-2}~{\rm sec}^{-1}$
(Piran et. al. 2001).   The X-ray flux reflects $E_k$, the
kinetic energy of the relativistic ejecta  (averaged over an angular
scale $1/\gamma$ corresponding to $\gamma \sim 10$). Hence, in this burst
$E_k/E_\gamma$ is larger by a factor of 2.25 than the average
value. This factor is similar to the energy fluctuations we find
in the patchy shell model for $p=2.2$ (see Fig. \ref{fig E}).
While in most GRBs that show a larger value of $E_k/E_\gamma$ we,
most likely, observe a $\gamma$-rays hot spot (Piran 2001).
According to this interpretation GRB 021004 is the first burst in
which a clear $\gamma$-ray cold spot has been seen.

JG thanks the Hebrew University for hospitality while this
research was done.  This work was partially supported by the
Horwitz foundation (EN) and by the  Institute for Advanced Study,
funds for natural sciences (JG).

\appendix{Appendix}

Because of the curvature of the afterglow shock, that is spherical
rather than planar, photons that are emitted from the shock front
at the same time in the source rest frame (i.e. at the same
radius), but at different angles from the line of sight, reach
the observer at different times. This causes two main effects:
first, the bulk of the energy that is emitted at a given time at
the source is delayed compared to a photon emitted on the line of
sight at that time, and second, at any given time the observer
receives photons that were emitted at different radii. In our
analysis we take the first effect (angular time delay) into
account (see Eq. \ref{eq tobs}), but the second effect
(angular smoothing) is neglected (see Nakar \& Piran 2003 for a solution
of the spherical symmetric afterglow light curve that takes a full
account of the angular effects). For spherical shells, the
angular smoothing produces an observed light curve which is a
smoothed version of the line of sight emission. The relative
importance of angular smoothing is determined by the ratio
$t_{\rm los}/t_{\rm ang}$, where
$t_{los}=\frac{1}{4c} \int^{R}_{0}dr/\gamma^{2}$
and $ t_{\rm ang} \approx R/4c\gamma^2$.
When the external density decays as a power law,
$n\propto r^{-k}$, the line of
sight time is: $t_{\rm los}=R/4(4-k)c\gamma^2$.
Most of the contribution to the observed flux at a time $t_{\rm obs}$, comes
from emission at radii $4\lesssim R/ct_{\rm obs}\gamma^2(R)\leq 4(4-k)$,
which correspond to
$t_{\rm los}\leq t_{\rm obs}\lesssim t_{\rm ang}$.
Hence, this effect is important when the light curve from the line of
sight varies significantly (compared to the smooth power law decay) on
time scales shorter than $t_{\rm ang}$
(i.e. $\Delta t_{\rm los}/t_{\rm los}<t_{\rm ang}/t_{\rm los}\sim 4$),
which corresponds to density variations on $\Delta R/R \lesssim 0.4$.
In such a case the observed light curve is significantly less variable
than the line of sight light curve.

We calculate the density profile
assuming that the observed (smoothed) light curve is similar to that
from the line of sight. Thus, the real density profile has to be more
variable than the one we obtain. The difference between the
two is smaller when the density profile increases with radius, and the
emission along the line of sight increases with time (compared to the
power law decay). In this case, the observed flux is dominated
by emission from large radii, near the line of sight (with a relatively
small contribution from large angles, for which the emission took place
at smaller radii, where the external density was relatively low) and
the angular smoothing effect is less important. However,
when the density drops, the angular effect becomes important.
Panaitescu \& Kumar (2000) have shown that even a sharp drop in the
density produces only a gradual temporal decay in the observed
light curve, and that the angular smoothing dictates a maximal
power law index of the temporal decay at late times.

The observed R-band light curve of GRB 021004 shows variations on
time scales of
$\Delta t/t\approx 1<t_{\rm ang}/t_{\rm los}\sim 4$,
therefore the angular smoothing effect is not negligible. This
effect can be seen by comparing our density profile to the
density profile obtained by Lazzati et al. 2002 (their Fig. 1),
which take the angular smoothing effect into account.
The main difference between the profiles is in the sharp density
drop after the first density bump (the density in the profile of
Lazzati et al. drops to one order of magnitude below the initial
density, while our profile drops back to the initial density).
This sharp drop is required, when the angular smoothing is
considered, in order to obtain the steep temporal decay in the
light curve after the first bump ($t\approx 10^{4}\;$sec).

This last result is obtained under the assumption of a full
spherical symmetry. However, it is more likely that the overdense
regions are concentrated in clumps and not in spherical shells
(see section 3.1 and Lazzati et al. 2002). The radial size of the
first clump is $\sim 10^{16}\;{\rm cm}\sim R_{i}/\gamma_{i}$,
where $R_{i}$ and $\gamma _{i}$ are the radius and
Lorentz factor of the blast-wave when it first interacts with the
clump. If we assume that the clump is spherical then its angular
size is $\sim 1/\gamma _{i}$ and at the beginning of the first
bump ($t\approx 10^{3}\;$sec) it ``fills'' most of the
observed region (the region within an angle of $1/\gamma$ around
the line of sight).
As the dominant emission is from the hot clump, the angular time
remains constant, $R/4\gamma^{2}_{i}$, while $t_{\rm los}$
grows. Therefore, the angular smoothing effect becomes less and
less important, and the approximation which neglects this effect
holds better than for a spherically symmetric external density
profile. Therefore, in this scenario, our method  yields  a good
approximation for the actual
density profile at radius $R$, averaged over an angle of $\sim 1/\gamma$
around the line of sight.

\end{document}